\documentstyle[twocolumn,aps,psfig]{revtex}
\tighten
\begin{document}
\draft

\title{Simulation of Flux Lines with Columnar Pins: \\
Bose Glass and Entangled Liquids}

\author{Prasenjit Sen$^{1}$, Nandini Trivedi$^{1\ast}$ and 
D.~M. Ceperley$^{2}$ }

\address{$^1$ Department of Theoretical Physics,
Tata Institute of Fundamental Research, 
Homi Bhabha Road, Mumbai 400005, India \\
$^2$Department of Physics and 
National Center for Supercomputing Applications, University of
Illinois at Urbana-Champaign, Urbana, Illinois, 61801, USA.}
\address{
\begin{minipage}[t]{6.0in}
\begin{abstract}
Using path integral Monte Carlo we simulate 
a 3D system of up to 1000 magnetic flux lines by 
mapping it onto a system of interacting bosons in (2+1)D. 
With increasing temperature we find a first
order melting 
of flux lines from an ordered solid to an entangled
liquid signalled by a finite entropy jump 
and sharp discontinuities in the defect density and the structure
factor $S({\bf G})$ at the first reciprocal lattice vector.
In the presence of a small number of strong columnar pins,
we find that the crystal is transformed into a Bose glass phase 
with patches of crystalline order nucleated around the trapped vortices
but with 
no overall positional or orientational order. This glassy phase
melts into a defected entangled liquid through a continuous transition.
\end{abstract}
\end{minipage}}
\date{\today}
\maketitle

\narrowtext

The seminal paper of Abrikosov in 1957~\cite{abrikosov}
showed, within a mean field calculation,
that for fields below a critical field $H_{c2}(T)$
a type II superconductor allows magnetic
flux to penetrate 
in the form of a triangular array of vortices each carrying a 
unit flux quantum $\phi_0=hc/2e$. This picture is modified considerably in 
the case of high $T_c$ cuprates which are quasi two-dimensional and 
have high transition temperatures~\cite{blatter_review}.
The most dramatic consequence
of the enhanced thermal fluctuations 
on the Abrikosov lattice 
is its melting 
into a vortex liquid by a first order transition
as indicated in resistivity,
neutron scattering,
specific heat, SQUID magnetometry  
and local Hall probe~\cite{expts_melting} measurements. 
Peak effect~\cite{tifr} measurements on NbSe$_2$ have also found
a reentrant phase diagram in the H-T plane~\cite{nelson,magro}.
Theoretical calculations have primarily been confined to 
density functional 
approaches on the layered system of pancake 
vortices~\cite{menon} valid for BSCCO, lattice simulations of 
frustrated 3D XY-models~\cite{3dxy}, Monte Carlo or Langevin dynamics
of the vortices~\cite{langevin},  
and determination of the melting line using Lindemann
criterion~\cite{houghton}.
Simulations of vortices, treated as line or 
polymer-like
objects, have been done using an analogy with interacting 
bosons~\cite{nelson} by variational and path integral techniques
~\cite{magro,nordborg}. 

The effect of pins on the 
structure and dynamics of the vortex lattice adds 
another dimension to vortex behavior.
The original work of Larkin and
Ovchinnikov~\cite{larkin} predicted that for $d <  4$, point pins destroy the 
long range order of the vortex lattice and break it up into small 
ordered coherence volumes.
From the point of view of practical applications where the aim is to minimize
the motion 
of vortices in order to reduce dissipation,
it is found that columnar pins along the vortex direction 
are extremely effective in pinning vortices~\cite{columnar}.
There have been very few simulations attempting to study the phases and phase
transitions in the presence of columnar pins. The ones based on
3D XY models assume an 
underlying numerical grid which affects the vortex structures 
for the small system sizes and high vortex densities
that have been studied~\cite{lee_stroud,3dxy}.

With this motivation, in this Letter we present the first simulations of
the structure and thermodynamics 
of up to 1000 vortices with and without strong columnar pins
using
a {\it continuum} path integral quantum Monte Carlo (PIMC) simulation of the
corresponding interacting bosons in (2+1)D.
There is no bias 
of a superimposed lattice and no 
prior input of the nature of wave functions or the type of the transition.
We extend previous simulations~\cite{nordborg} to include columnar pins
on considerably larger lattices.

Our main findings are as follows--
In the pure system, we see the reentrant nature of the melting line in agreement with the
predictions of Nelson~\cite{nelson}.
We find a triangular
solid phase at low temperatures $T$
which melts into an {\it entangled} liquid phase at higher $T$
by a first order transition with a finite
entropy jump. 
The defect density, and 
structure factor at the first
reciprocal lattice vector, $S({\bf G})$, 
also show sharp discontinuities at the transition.
In the solid phase,
$S({\bf G})\sim {\cal O}(N)$ and drops
sharply to ${\cal O}(1)$ in the liquid phase. 
In the presence of 10\% columnar pins, 
the vortex lattice transforms into a Bose glass phase at low $T$. 
The Bose glass consists of 
patches of ordered regions with only short range
positional {\em and} orientational order.
This phase is qualitatively different from the Bragg glass phase
with point pins\cite{giam}.
With increasing temperature, there is a transition from a Bose glass to an 
entangled `defected liquid' 
which is considerably smeared.

Consider a system of $N$ flux lines in 3D 
characterized by a bending energy
$\tilde{\epsilon_1}
= \lbrace \phi_0/(4\pi \lambda) \rbrace^2 $~\cite{note} 
where $\lambda$ is the London penetration depth. 
The inter-vortex interaction is approximated as 
$V(r_{ij}) = \epsilon_0 K_0( {r_{ij}}/{\lambda} ) $ 
where $\epsilon_0 = \phi_0^2/({8\pi^2 \lambda^2})$ is the scale of the 
inter-vortex potential, 
$K_0$ is the modified Bessel
function of the first kind and ${\bf r}_i(z)$'s are the 2D 
position vectors of the flux lines in a
plane perpendicular to their length at a height $z$. 
$K_0(r)\sim -{\rm ln}{(r)}$ as $r \rightarrow 0$ 
and is valid in the high vortex
density limit, the regime studied in Ref.~\cite{nordborg}. 
In this paper, we fix the density on the
lower part of the reentrant melting curve~\cite{magro} and therefore retain the 
full $K_0(r)$ form of the potential.

The partition function of the flux lines 
is mapped onto the world lines or Feynman paths
of bosons~\cite{nelson} 
described by the Hamiltonian
\begin{equation}
H = -\Lambda^2 \sum_{i=1}^N \nabla_i^2+ \sum_{i<j} K_0(r_{ij})
\label{eq:hamil}
\end{equation}
The bosonic path integrals are in 
two spatial dimensions 
and one imaginary time direction, 
which is equivalent to the thickness of the sample 
along the direction of the magnetic field.
All lengths are measured in units of $\lambda$ and energies in units of 
$\epsilon_0$.
The dimensionless de Boer parameter 
$\Lambda={k_B T}/({2 \tilde{\epsilon_1} 
\lambda^2 \epsilon_0})^{1/2}$
is a measure of quantum fluctuations (the relative strength of 
the kinetic energy {\it vs} the interaction energy) of the boson
system, or equivalently a measure of temperature $T$ in the vortex
system.
Once the mapping is established, 
we do a 
PIMC simulation~\cite{ceperley} 
of interacting bosons
at an inverse boson temperature $\beta=1/T_b$~\cite{parameters}.
Periodic boundary conditions are used in
all directions and permutations between bosons are included\cite{bosons}.
If $T_b < \Lambda^2 \rho$,
where $\rho=N \lambda^2/({\rm Area})$ 
is the dimensionless density of the bosons,
exchanges between the particles are possible 
and can generate entangled vortex lines.

In Fig.~\ref{fig:picture}(a) 
we show the structure of 1000 vortices at low 
temperature $\Lambda=0.045$ 
where the vortex system shows a beautiful triangular
lattice with small excursions of the vortex lines about their equilibrium
positions seen in Fig.~\ref{fig:pathplot}(a). For $\Lambda>~ 0.064$
the vortices melt into an entangled liquid as seen from the actual
configurations of the vortex lines in Fig.~\ref{fig:pathplot}(b). 
Concomitantly, the superfluid density $\rho_s$ 
of the corresponding boson system
jumps from zero in the solid phase to a finite value in the liquid phase.
The entanglement correlation length~\cite{nelson}, which is the typical length
along the $z$-direction for vortex lines to cross, 
is on the order of the sample size in the solid and drops 
to $\sim 4a_0$, where $a_0$ is the planar lattice constant,
in the liquid phase.

The degree of positional order is quantified by 
the structure factor
$S({\bf k}={\bf G})$ at the first
reciprocal lattice vector of the triangular lattice.
As seen in Fig.~\ref{fig:sk}, with increasing temperature there is a sharp 
transition from a crystalline to a liquid phase at $\Lambda_m \sim 0.064$.
In a temperature-driven first order transition, the free energy is 
continuous across the transition but the internal energy and entropy 
are discontinuous. In analogy in the mapped boson problem at $T=0$, 
the total energy is continuous but there are sharp jumps in the potential
and kinetic energies 
between $\Lambda=0.062$ and $0.064$. 
The melting transition is located by 
the intersection of the polynomial
fits to the energy in the liquid and solid phases at
$\Lambda_m = 0.064$ in good agreement with Ref.~\cite{magro}. 

In the
crystal $S({\bf k})$ shows Bragg peaks
with $S({\bf G})\sim {\cal O} (N)$, where
$N$ is the number of vortices, while above $\Lambda_m$, 
$S({\bf k})$ is liquid-like with
$S({\bf G})\sim {\cal O} (1)$.  
By fitting the structure factor data to the form
$S({\bf G}) = 1 + (N-1)\exp{(-G^2\langle u^2\rangle)}$ for a finite
system we have calculated the 
Lindemann number $c_L = {\langle u^2 \rangle}^{1/2}/a_0=
0.24$
at the transition just within the solid phase, in good
agreement with other works\cite{blatter_review},  
where $\langle u^2 \rangle$
is the mean squared displacement of the particles from their lattice
positions~\cite{draeger}. 

In addition to the positional correlations, the orientational order 
for six-fold symmetry is described by 
$\Psi_6({\bf r}_i) = (1/6)\sum_j e^{i6\theta_{ij}}$,
where the sum over $j$ runs over all the nearest neighbors of $i$ in a
triangulation picture, and $\theta_{ij}$ is the angle the bond between
particles at ${\bf r}_i$ and ${\bf r}_j$ 
makes with an arbitrary reference axis. The
orientational correlation function defined as
$g_6(r) = \langle \Psi_6(r)\Psi_6^*(0) \rangle$
also shows long range order in the clean system at low $T$. 
We find that the correlation lengths for positional and orientational order 
can reliably
be extracted only from 
data on large system sizes $\sim 1000$ vortices and we find that both
types of order vanish at $\Lambda_m$.

A measure of imperfections in the triangular arrangement of the flux
lines is the number of topologically-defected non six-fold coordinated
sites in the Delaunay triangulation\cite{preparata}
of imaginary time slices of the system.
This defect density is small at low temperatures in the clean system and
shows a sharp jump across the melting transition (Fig.~\ref{fig:defect}).

\noindent {\em Effect of Columnar Pins:}

In the boson picture, randomly placed potential wells 
are correlated in the imaginary-time direction and correspond
to columnar pins in the vortex representation. 
We model the columnar pin as an attractive delta function
which traps a vortex all along its length. In the boson picture
the trapped vortex corresponds to a fixed classical particle.
The vortex phase in the presence of columnar pins has been previously 
referred to as a Bose glass phase~\cite{nelson}. 
Our simulations provide the first detailed
description of the structure and properties of this so-called Bose glass.
In Fig.~\ref{fig:picture}(b)
we show the configuration of the 1000 vortex system at $\Lambda=0.045$ 
with $10\%$ of the vortices, those trapped by columnar pins, held fixed.
The pins are found to seed or nucleate an ordered crystalline patch, 
however, since the pins are randomly located, the different 
patch orientations are not commensurate and frustrates the 
translational and orientational order of the lattice.
Columnar pinning dramatically reduces $S({\bf G})$
from around 250 in the pure 1000 vortex system
to about 6 in the presence of pins at $\Lambda=0.045$ in the Bose glass
phase (Fig.~\ref{fig:glass}(a)). In addition, $S(k)$ has 
a jagged peak, 
with shoulders and split peaks, which are characteristic of glasses.
The Bose glass is rather different from a structural glass 
or a frozen liquid which has a 
peak height of $S({\bf G})\approx 1.5$. 
At higher temperatures the Bose glass melts into a 
defected liquid phase where the unpinned 
vortices are entangled (Fig.~\ref{fig:pathplot}(c)).
In the parameter region investigated, we do not find any evidence for a 
disentangled liquid either with or without pins.
The Bose glass phase that we find in these simulations is in a different limit
from that studied by Nelson 
and Vinokur~\cite{nelson-vinokur} 
where the number of columnar pins is much larger than the number of vortices.

The topological defects generated by the columnar pins in the vortex lattice 
destroy 
the positional and orientational order within few lattice 
spacings (Fig.~\ref{fig:glass}(b)) in contrast
to a Bragg glass~\cite{giam} in the presence of weak point pins which
retains long range orientational order and algebraic positional order.

Columnar pinning has a marked effect on the first order melting transition.
The kinetic and potential energies vary smoothly through the transition.
The sharp drop in $S({\bf G})$ in Fig.~\ref{fig:sk} also disappears. 
The
sharp jump in defect density is replaced by a much smoother 
increase through the transition as seen in Fig.~\ref{fig:defect}.
The local variations in the melting temperature caused by the quenched
random potential below a certain critical dimension
can cause a sharp first order transition to
get rounded off in presence of quenched randomness~\cite{imrywortis}. 
It is possible that at lower pin densities, the first order jumps in
various quantities persist with reduced magnitudes. Further simulations
are in progress to study the evolution of the sharp first order
transition with increasing defect density.

P. Sen would like to acknowledge B. Militzer, G. Bauer and E. Draeger
for help with the code.
We would also like to thank
S. Banerjee, S. Bhattacharya, C. Dasgupta, T. Giamarchi, A. Grover, 
G. Menon, and A. Paramekanti for useful discussions.
We acknowledge the support of the NCSA,
UIUC for computational resources and software
and the NSF grant no. DMR 98-02373.

\noindent $^\ast$ e-mail: ntrivedi@tifr.res.in

{\centerline {\bf FIGURE CAPTIONS}}
\begin{figure}
\caption{
Time averaged locations of 
1000 vortices at $\Lambda=0.045$ 
on a slice 
perpendicular to the z-direction and then averaged over many slices
along the thickness of the vortex system;
red regions represent higher vortex density and blue regions are lower
vortex density.
(a) Triangular lattice of vortices in the pure system. 
(b) Frozen Bose glass phase with
10\% strong columnar pins
for one realization of disorder. A vortex, 
with a delta function density distribution, 
is trapped at the center of a pin 
(not shown). The blue region is the region of reduced vortex density 
from which other vortices are repelled by the trapped vortex.
}
\label{fig:picture}
\end{figure}

\begin{figure}
\caption{A snapshot of vortex positions on the top layer
(shown as black dots) and the projections on to the top layer 
of the vortex line (shown in red). 
Only a portion of the 1000 vortex system is shown for clarity. 
(a)For $\Lambda=0.045$ in the solid phase there are very 
small excursions of the vortex lines.
(b)For $\Lambda=0.065$ just above the melting transition in the
clean system the vortex lines are in a highly entangled liquid phase
and the vortex density upon averaging is uniform.
(c)For $\Lambda=0.065$ with 10\% pins the system is also melted
into a defected entangled liquid. In the portion shown there are 3 pins 
each trapping one vortex, 
(isolated black dots), 
which excludes other vortices around it.
}
\label{fig:pathplot}
\end{figure}

\begin{figure}
\caption{The structure factor $S({\bf k}={\bf G})$ at the first
reciprocal lattice vector of the triangular lattice as a 
function of $\Lambda$ for 100 vortices at a density $\rho=0.02$.
The circles show
$S({\bf G})$ in the clean system which shows a transition
from a crystalline to a liquid phase at $\Lambda\sim 0.064$. 
The squares show
$S({\bf G})$ with 10\% columnar pins which shows a much reduced
degree of translational order in the Bose glass phase.
}
\label{fig:sk}
\end{figure}

\begin{figure}
\caption{
Fraction of non six-fold coordinated sites in the clean and pinned
systems. In the clean system the defect density shows a sharp jump across the
transition at $\Lambda=0.064$. 
In the presence of pins, the topological defect density is higher 
within the phases and the transition is smoothed out.
}
\label{fig:defect}
\end{figure}

\begin{figure}
\caption{
A 1000 vortex system with 10\% columnar pins at $\Lambda=0.045$ in the
Bose glass phase. 
(a) The structure factor $S(k)$ {\it vs.} $k$ showing a much reduced peak
from about 250 in the crystal to $\sim 6$ in the Bose glass phase; but higher 
than in the liquid with peak height $~1.5$.
(b) The radial positional distribution function $g(r)$ and
the orientational correlation function $g_6(r)$. An envelope of the form
$\exp(r/\xi)$ can be fit for both
$g(r)$ and $g_6(r)$ with $\xi=13 \lambda$ and $\xi=15 \lambda$ respectively. 
For density $\rho=0.02$, the inter-vortex separation $a_0=7.6\lambda$ 
so the decay length of the correlations is $\sim 2a_0$
in the Bose glass phase.
}
\label{fig:glass}
\end{figure}


\begin{thebibliography}{unsrt}
\bibitem{parameters}
We use $\beta=1/T_b=20000$ 
which is discretized into $M=128$ slices. 
The time step $\delta \tau \equiv \beta / M = 156.25$ 
and the typical energy $E\sim 10^{-3}$, which ensures that
the Trotter error $\delta\tau E \ll 1$.
\bibitem{bosons} It is possible to have an entangled liquid with either 
bosons or boltzmannons, but
bosons generate the $T_{b}=0$ state at a higher temperature 
and are thus computationally more
convenient\cite{ceperley}.
\end{thebibliography}
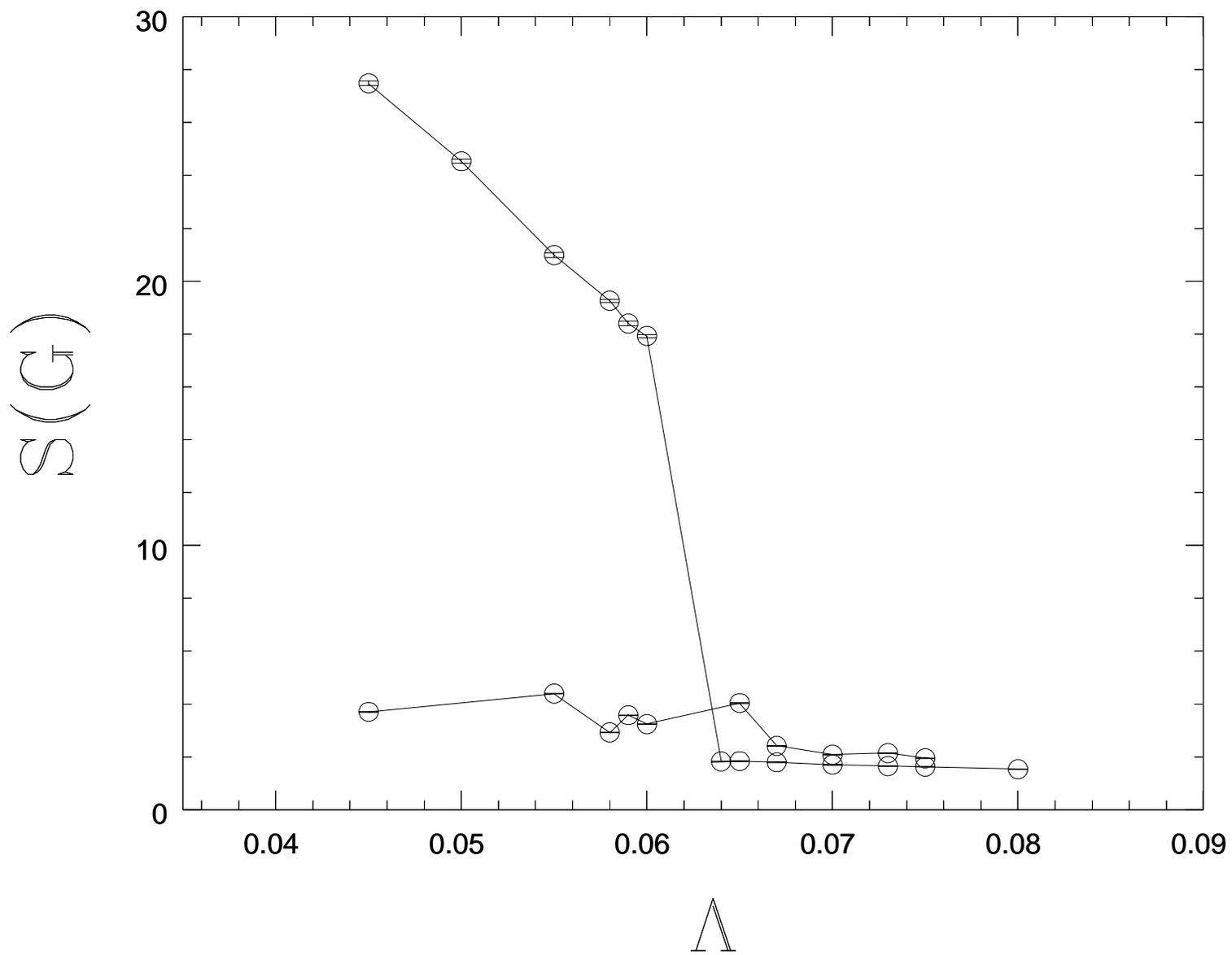
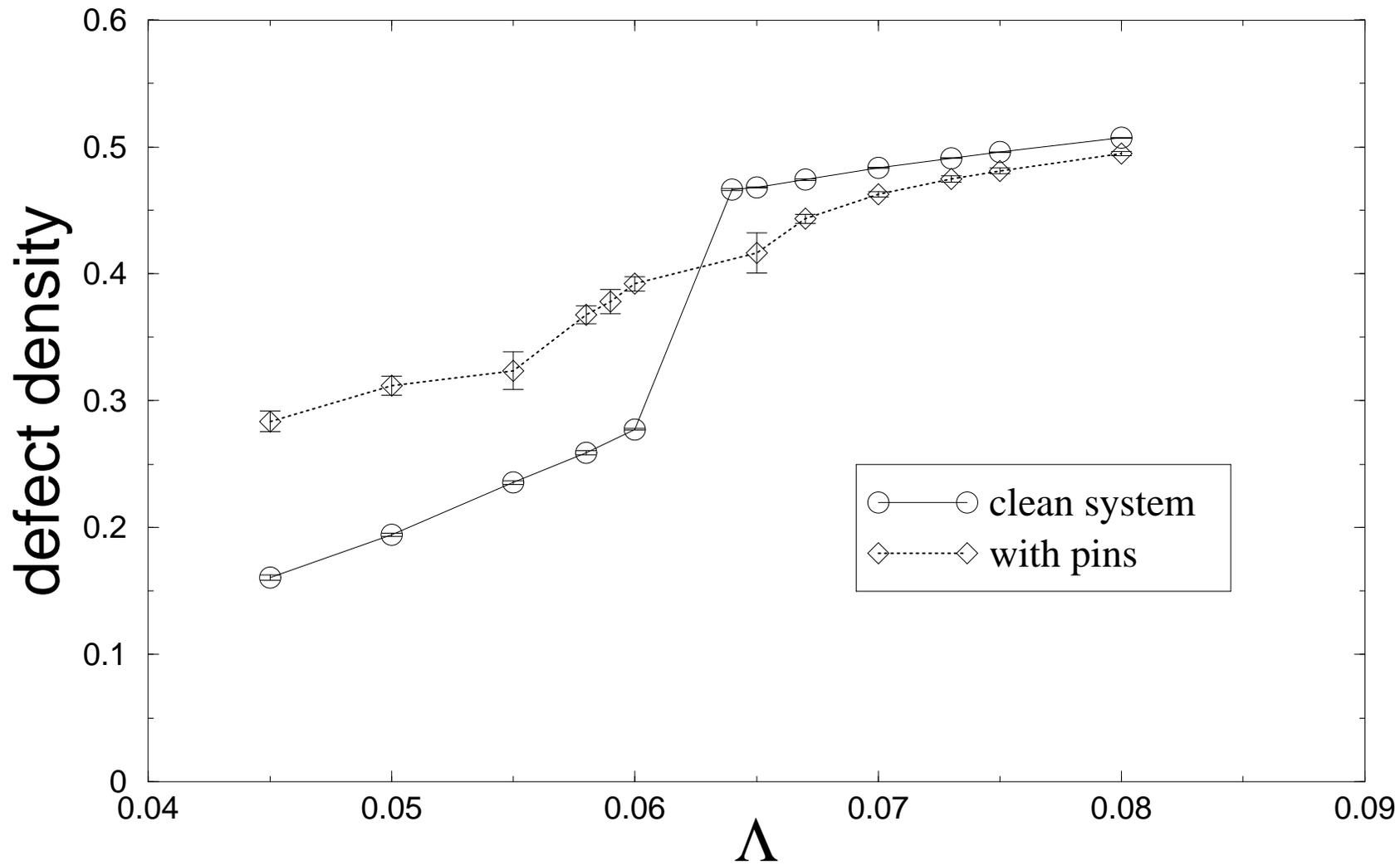
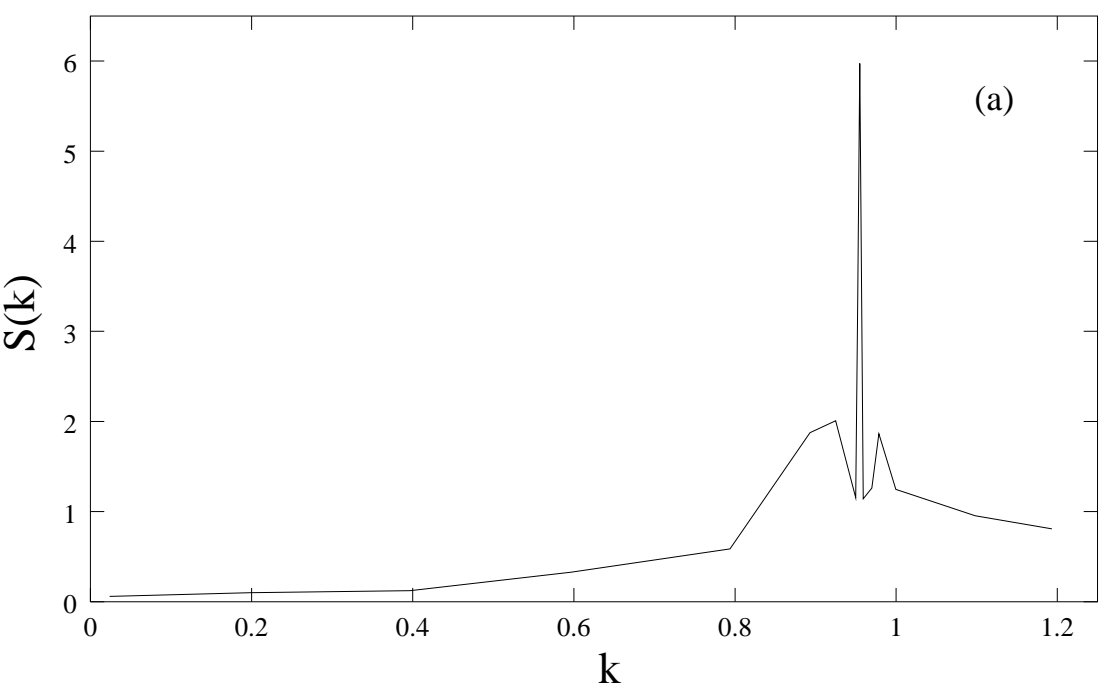
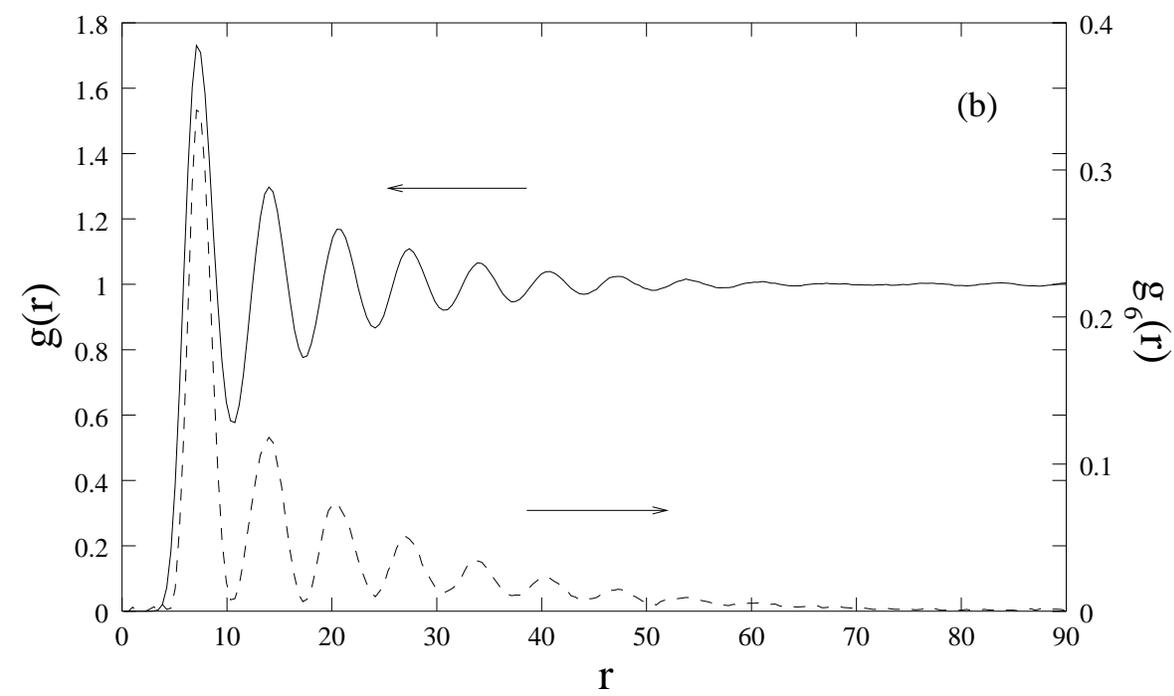
\end{document}